\documentclass
{modified_aipproc}

\usepackage{amssymb}
\layoutstyle{8x11double}

\begin{document}


\title{Measuring Dark Matter at a Collider\footnote{Plenary talk given at PASCOS05, Gyeongju, Republic of Korea, June 2005.}}


\author{Andreas Birkedal}{
  address={Santa Cruz Institute for Particle Physics, Santa Cruz, CA 95064, USA \\ Institute for Fundamental Theory, University of Florida, Gainesville, FL 32611, USA}}

\begin{abstract}
We investigate the need and prospects for measuring dark matter properties at particle collider experiments.  We discuss the connections between the inferred properties of particle dark matter and the physics that is expected to be uncovered by the Large Hadron Collider (LHC) and the International Linear Collider (ILC) and motivate the necessity of measuring detailed dark matter properties at a collider.  We then investigate a model-independent signature of dark matter at a collider and discuss its observability.  We next examine the prospects for making precise measurements of dark matter properties using two example points in minimal supergravity (mSUGRA) parameter space.  One of the primary difficulties encountered in such measurements is lack of constraint on the masses of unobservable heavy states.  We discuss a new method for experimentally deriving estimates for such heavy masses and then conclude.
\end{abstract}

\maketitle


\section{Introduction}

The existence of non-baryonic, non-luminous matter appears now to be firmly established.  The amount of such 'dark' matter has been measured at the 10\% level.  The current best limit is given in terms of $\Omega_{\chi}$, the fractional energy density of the universe that is contained in dark matter, and $h$, the reduced Hubble parameter~\cite{Spergel:2003cb}:

\begin{equation}
0.094 \leq \Omega_{\chi} h^2 \leq 0.129 \ \  {\rm (at\ 2 \sigma)}.
\label{wmap}
\end{equation}

The nature of dark matter is one of the greatest unsolved mysteries of science.  Usually, one hypothesizes an extra stable electrically neutral particle that can serve as dark matter.  Most particle physics models that attempt to explain the weak scale and electroweak symmetry breaking contain such additional particles among their non-Standard Model particle spectra.  Since one of these models might be discovered at the LHC or ILC, it is interesting to ask what, if anything, such colliders can tell us about dark matter.  This talk focuses specifically on answers to this question.

\section{Connections Between Dark Matter and Colliders}

\begin{figure}
  \includegraphics[width=75mm]{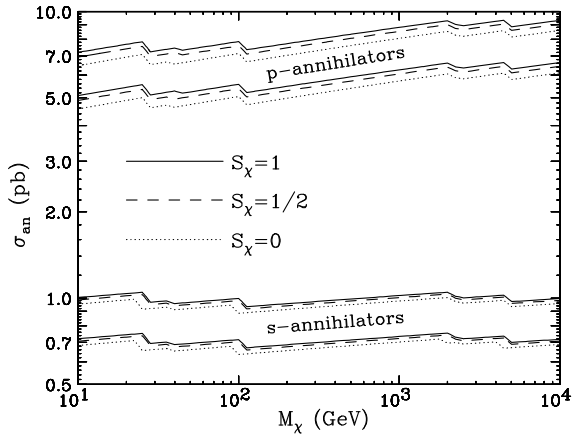}
  \caption{Values of the quantity $\sigma_{an}$ allowed at $2\sigma$ level as a function of the dark matter mass.  The lower (upper) band is for models where $s$-wave ($p$-wave) annihilation dominates.  Taken from Ref.~\cite{Birkedal:2004xn}.}
\label{fig:sandp}
\end{figure}

Precious little is known about dark matter, except for its gross density, given in Eqn.~\ref{wmap}.  If one assumes that dark matter comes from a thermal relic, this gross density is, in large part, determined by the dark matter annihilation cross section: $ \sigma \left( \chi \, \chi \rightarrow SM \, SM \right)$.  In fact, the present-day dark matter abundance is roughly inversely proportional to the thermally averaged annihilation cross section times velocity\footnote{Usually the cross section times velocity can be conveniently expanded in powers of relative dark matter particle velocity: $\sigma v = \sum_{J} \sigma_{an}^{\left(J\right)} v^{\left(2J\right)}$.  Commonly only the lowest order non-negligible power of $v$ dominates.  For $J=0$, such dark matter particles are called $s$-annihilators, and for $J=1$, they are called $p$-annihilators; powers of $J$ larger than $1$ are rarely needed.}: $\Omega_{\chi} h^2 \propto 1/\langle \sigma v \rangle$.  Figure~\ref{fig:sandp} shows the constraint on the annihilation cross section as a function of dark matter mass that results from Eqn.~\ref{wmap}.  Remarkably, this constraint is effectively independent of dark matter mass\footnote{This effect is due to the changing number of degrees of freedom at the time of freeze-out as the dark matter mass is changed.}, and points to cross sections expected from weak-scale interactions (around $0.8\, pb$ for $s$-annihilators and $6\, pb$ for $p$-annihilators).  It points to the possibility that the solution to the dark matter puzzle is connected to the explanation for the weak scale, leading to the idea of Weakly Interacting Massive Particle (WIMP) dark matter.  Such WIMPs exist not only in supersymmetric theories, but also in theories involving extra dimensions~\cite{KKDM} and 'little Higgs' theories~\cite{LHDM} as well.  The LHC and the ILC are specifically designed to probe the origin of the weak scale, so dark matter and future collider physics appear to be intimately connected.

Many experiments exist that hope to discover dark matter at many distances from earth.  However, they, by themselves, are not able to solve the dark matter puzzle.  Direct detection experiments hope to measure the mass of dark matter particles and their scattering cross section off atomic nuclei, $\sigma\left(\chi N \rightarrow \chi N\right)$.  Since any measured rate of scattering depends on the number of dark matter particles present, astrophysical uncertainties in halo models affect the accuracy of measurements.  Furthermore, to claim that the dark matter puzzle has been solved, we must determine that a given particle (or set of particles) can provide the correct $\langle \sigma v \rangle$ shown in Fig.~\ref{fig:sandp}.  It is generically not possible to relate $\sigma\left(\chi N \rightarrow \chi N\right)$ to the required $\sigma\left(\chi \, \chi \rightarrow SM \, SM\right)$.  Indirect detection experiments attempt to discover dark matter through its present-day annihilation in regions of high density such as the galactic center.  Again, significant astrophysical uncertainties exist.  Also, these experiments hope to measure either a single annihilation final state, such as  $\sigma\left(\chi \, \chi \rightarrow \nu \, \nu\right)$, or a subset of final states, such as those with charged particles in the final state.  While these final states do contribute to $\sigma\left(\chi \, \chi \rightarrow SM \, SM\right)$, again, there is not a model-independent relation between the two.  As a result, it is necessary to make detailed measurements of the particles and couplings involved in dark matter annihilation.  These detailed measurements can only be done at particle collider experiments, such as the LHC and ILC.

\begin{figure}[t]
\begin{tabular}{cc}
  \includegraphics[width=75mm]{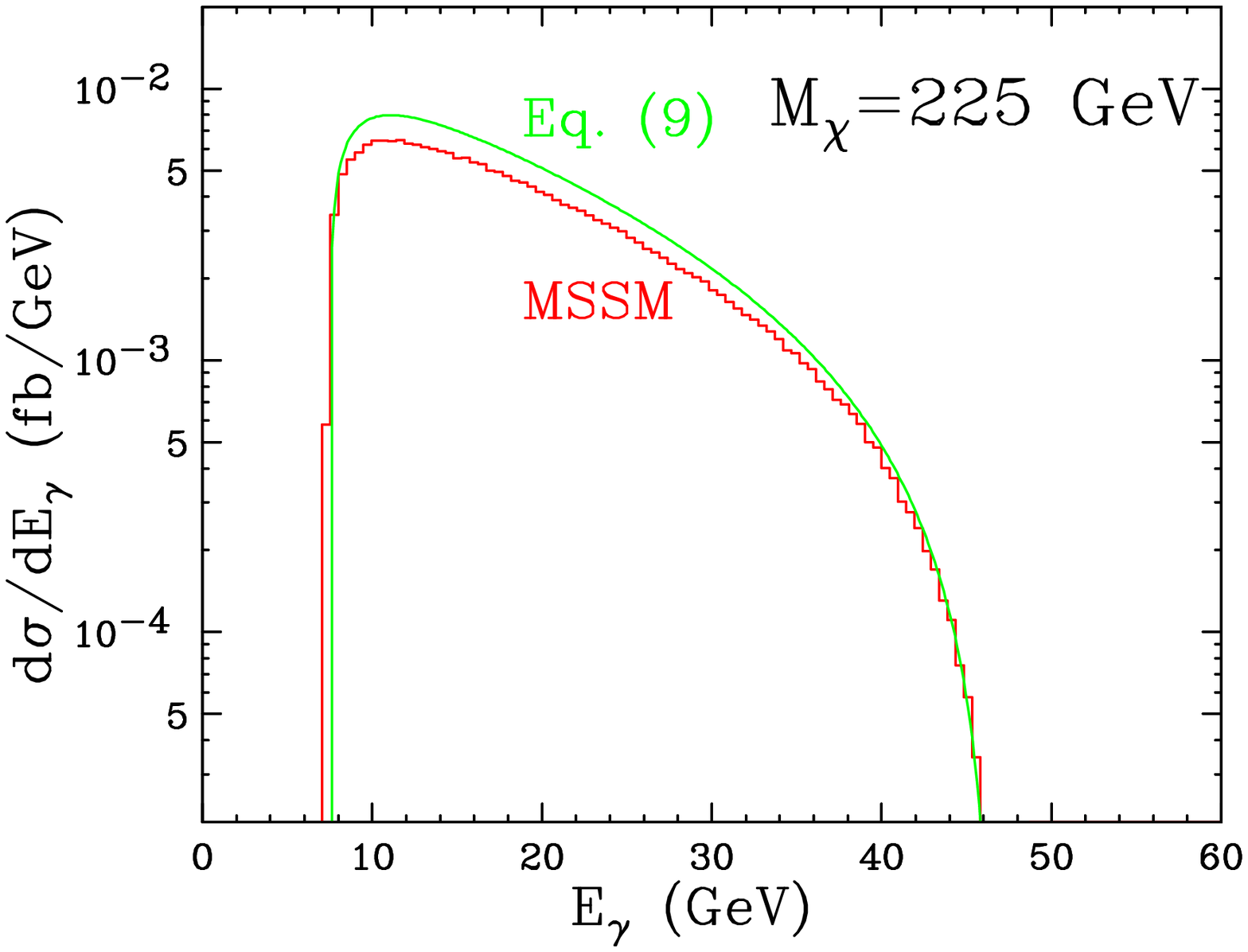}&
  \includegraphics[width=75mm]{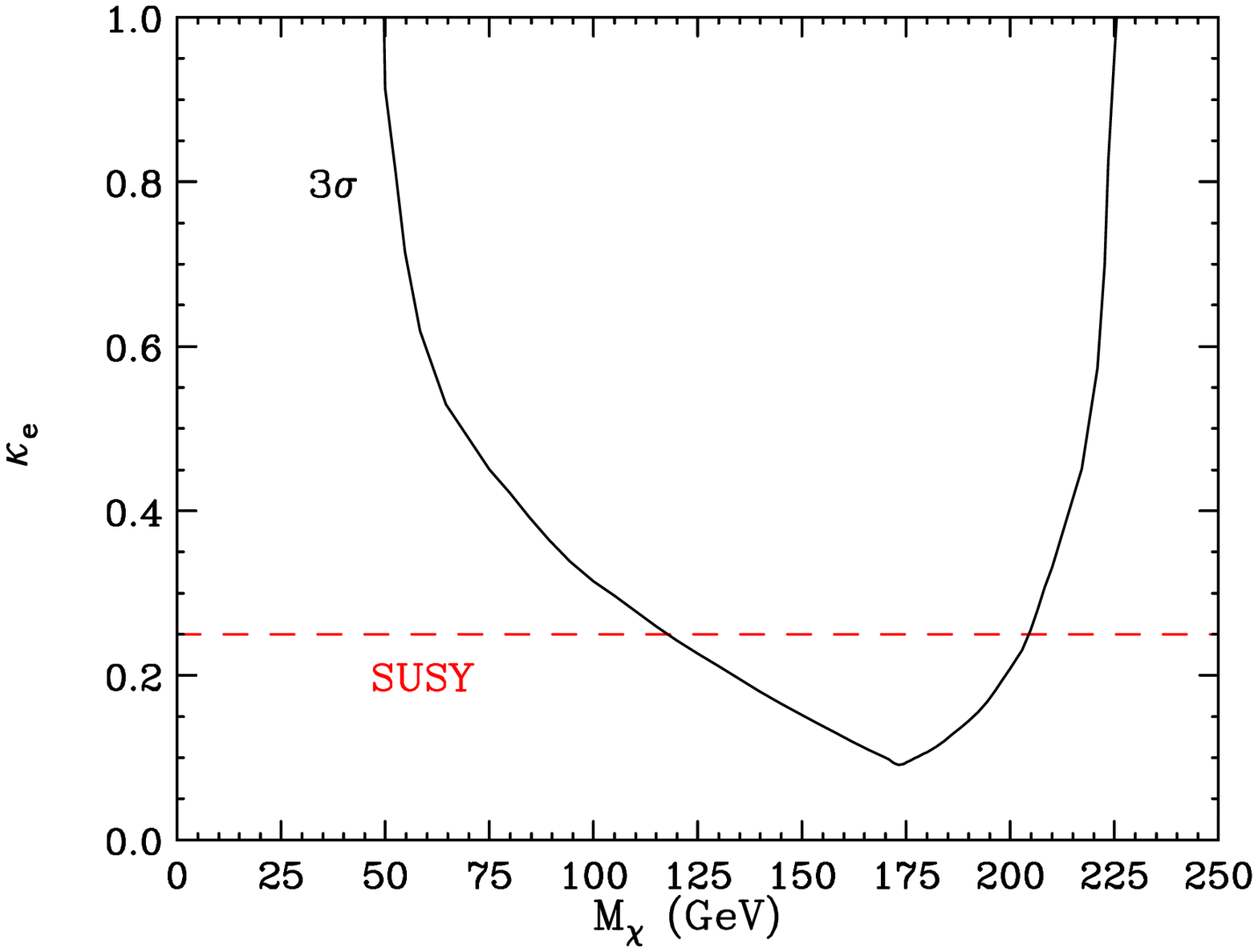}
\end{tabular}
  \caption{{\it Left panel:} Comparison between the photon spectra from the process 
$e^+e^-\to2\chi_1^0+\gamma$ in the explicit supersymmetric models defined in Ref.\cite{Birkedal:2004xn} (red/dark-gray) and the spectra predicted by Eqn.~\ref{rate} for a 
$p$-annihilator of the corresponding mass and $\kappa_e$ 
(green/light-gray). Taken from Ref.~\cite{Birkedal:2004xn}.  {\it Right panel:} The reach of a 500 GeV unpolarized electron-positron collider with 
an integrated luminosity of 500 fb$^{-1}$ for the discovery of 
$p$-annihilator WIMPs, as a function of the WIMP mass $M_\chi$ and the 
$e^+e^-$ annihilation fraction $\kappa_e$. The 3 $\sigma$ (black) contour is shown, along with an indication of values one might expect from supersymmetric models (red dashed line, labelled 'SUSY'). Only statistical uncertainty is included.}
\label{carrots}
\end{figure}

\section{Model-Independent Collider Search for Dark Matter}

Since the WIMP hypothesis and the known abundance of dark matter give specific values for the dark matter annihilation cross section, one might hope this cross section can be turned into a rate for a measurable process at a particle collider.  Such a model-independent collider signature does indeed exist~\cite{Birkedal:2004xn}\footnote{Due to space restrictions, here we are skipping many intermediate steps.  For further detail, we refer the reader to the original paper, Ref.~\cite{Birkedal:2004xn}.} and has implications for indirect detection experiments~\cite{Birkedal:2005ep}.

To arrive at the model-independent signature, we start with cosmological data, namely the allowed values for $\sigma_{an}$ from Fig.~\ref{wmap}.  We introduce the parameter $\kappa_e$:

\begin{equation}
\kappa_e \equiv \sigma\left(\chi \chi \rightarrow e^+ e^-\right)/\sigma\left(\chi \, \chi \rightarrow SM \, SM\right)
\end{equation}

This equation relates the dark matter annihilation cross section to the cross section involving $e^+ e^-$ in the final state.  Then we use crossing symmetries to relate $\sigma\left(\chi \chi \rightarrow e^+ e^-\right)$ to $\sigma\left(e^+ e^- \rightarrow \chi \chi\right)$.  Finally, we use collinear factorization to relate $\sigma\left(e^+ e^- \rightarrow \chi \chi\right)$ to the process $\sigma\left(e^+ e^- \rightarrow \chi \chi\, \gamma\right)$.  We thus have gone from cosmological data, in the form of $\sigma_{an}$, to a prediction for the rate $e^+ e^- \rightarrow \chi \chi\, \gamma$:

\begin{eqnarray}
& &\hskip-.8cm
\frac{d \sigma}{dx\, d\cos\theta} (e^+e^-\to 2\chi+\gamma)\,\approx\,\frac{\alpha\kappa_e \sigma_{\rm an}}{16\pi} \frac{1+(1-x)^2}{x} \nonumber \\
& &\hskip-.8cm 
\times 
\frac{1}{\sin^2\theta}2^{2J_0} (2S_\chi+1)^2\,  
\left(1-\frac{4M_\chi^2}{(1-x)s}\right)^{1/2+J_0}.
\label{rate}
\end{eqnarray}

Here $x=2E_\gamma / \sqrt{s}$, $\theta$ is the angle between the photon and the incoming electron, $S_\chi$ is the spin of the WIMP, and $J_0$ is the dominant value of $J$ in the velocity expansion for $\sigma v$, defined earlier.

The accuracy of our method is illustrated in Figure~\ref{carrots}.  The left panel in Figure~\ref{carrots} shows how the formula in Eqn.~\ref{rate} (green line, labelled as 'Eq. (9)') compares with the exact calculation in a supersymmetric theory (red line, labelled 'MSSM') with a WIMP mass of $225$ GeV.  It is clear that the agreement is quite good, especially for the harder photons, which are most useful in our approach.  The right panel shows the expected reach in $\kappa_e$ for a $500$ GeV linear $e^+ e^-$ collider as a function of the WIMP mass.  We have included background from $e^+ e^- \rightarrow \nu \nu \gamma$.  We show the $3 \, \sigma$ contour (black solid line) including statistical errors and also indicate the values one might expect from supersymmetric models (the region below the red dashed line labelled 'SUSY').  It is clear that this signal, while challenging, could provide direct evidence of WIMP production at colliders.

\section{Testing Supersymmetric Cosmology at the ILC}

The above model-independent rate is rarely the dominant collider signature of new physics within a given model.  It therefore makes sense to look at more model-dependent processes; we choose the minimal supersymmetric extension of the Standard Model (MSSM).  The WIMP is taken to be the lightest supersymmetric particle (LSP), usually the lightest neutralino, $\tilde{\chi}^0_1$.  If the LSP is the dark matter particle, the generic collider signatures 
of supersymmetry all involve missing energy due to the two stable $\tilde\chi^0_1$s
escaping the detector. In order to prove that the missing 
energy particle is indeed a viable WIMP dark matter candidate, one needs to calculate
its expected present relic abundance. One thus needs to measure all parameters 
 which enter the calculation of $\langle \sigma v \rangle$. 
In the most general MSSM 
there are more than 100 input parameters at the weak scale, but fortunately, a 
lot of them are either tightly constrained or not very relevant for the dark matter calculation. Nevertheless, 
there are still several relevant parameters left; they must be
determined from collider data. Of course, the relevance of any one parameter
depends sensitively on the parameter space point.  We now look at two points: first, a point from the 'bulk region,' and then a point from the 'focus point region,'~\cite{Feng:1999mn} both from mSUGRA, a convenient subset of the larger MSSM~\cite{Birkedal:2004ALCPG,Gray:2005ci,Birkedal:2005jq}.

\begin{figure}
  \includegraphics[width=75mm]{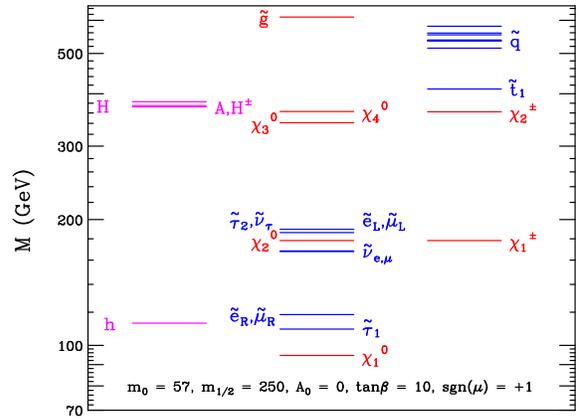}
  \caption{Mass spectrum of point B' from the bulk region.  Taken from Ref.~\cite{Birkedal:2005jq}.}
\label{fig:Bplevels}
\end{figure}

Our first point, point B' from the bulk region, is defined by the following
input parameters\footnote{Here we have used Isajet 7.69~\cite{Paige:2003mg} 
and DarkSUSY~\cite{Gondolo:2002tz} except for calculations where coannihilations are important, 
in which case we have used micrOMEGAS~\cite{Belanger:2001fz}.}: 
$m_0 = 57$ GeV, $m_{1/2} = 250$ GeV, $A_0 = 0$, $\tan \beta = 10$ and ${\rm sign}(\mu) = +1$.
Fig.~\ref{fig:Bplevels} shows that the particle spectrum at this point is quite light.  
The two lightest neutralinos, the lightest chargino and all of the sleptons have masses below $200$ GeV.  
All of the squarks are lighter than $600$ GeV.  The heaviest particle, the gluino, only weighs 
$611$ GeV. Therefore, one would expect colliders to have significant discovery and 
measurement capabilities.

\begin{figure}[t]
\begin{tabular}{ccc}
\includegraphics[width=50mm]{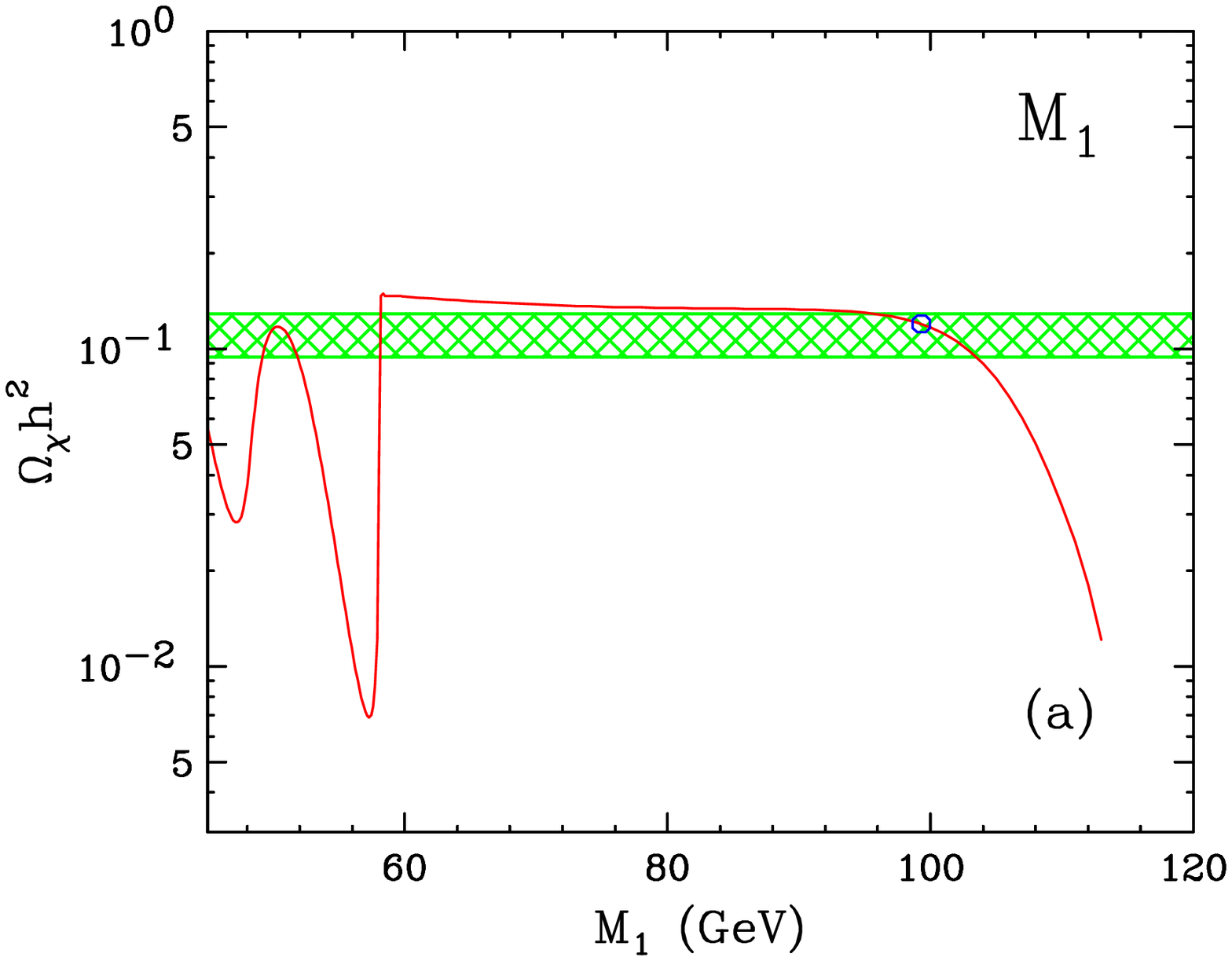}&
\includegraphics[width=50mm]{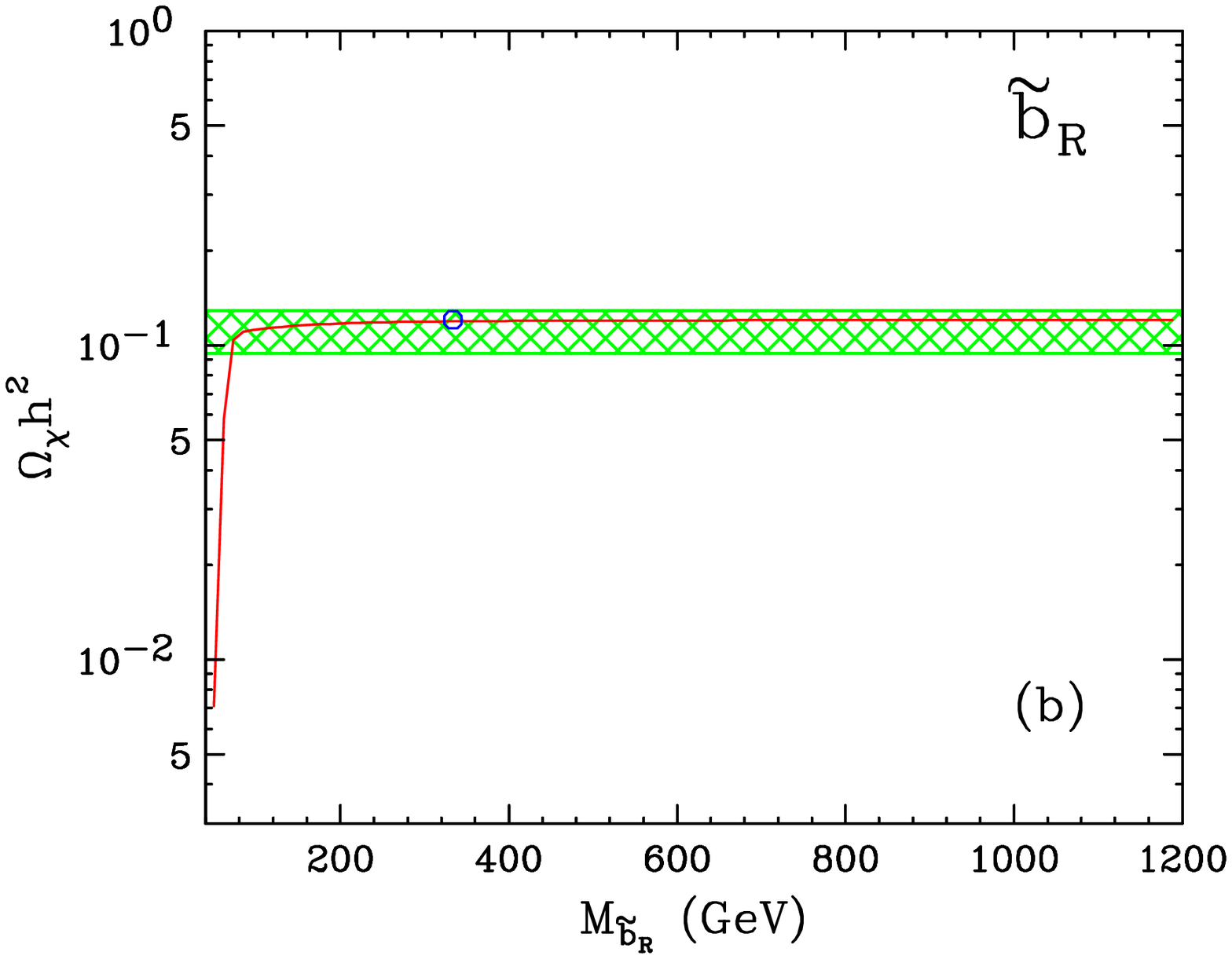}&
\includegraphics[width=50mm]{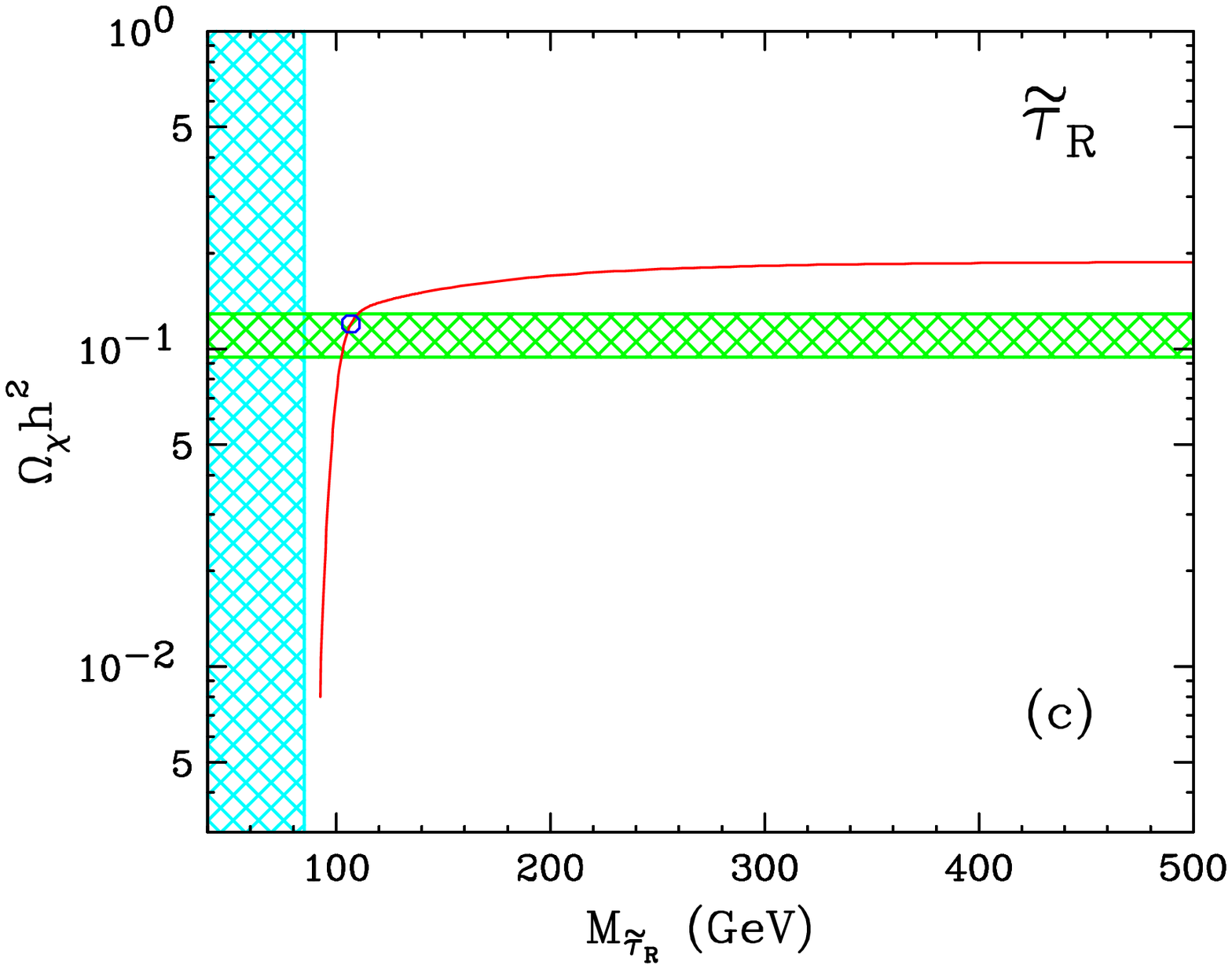}\\
\includegraphics[width=50mm]{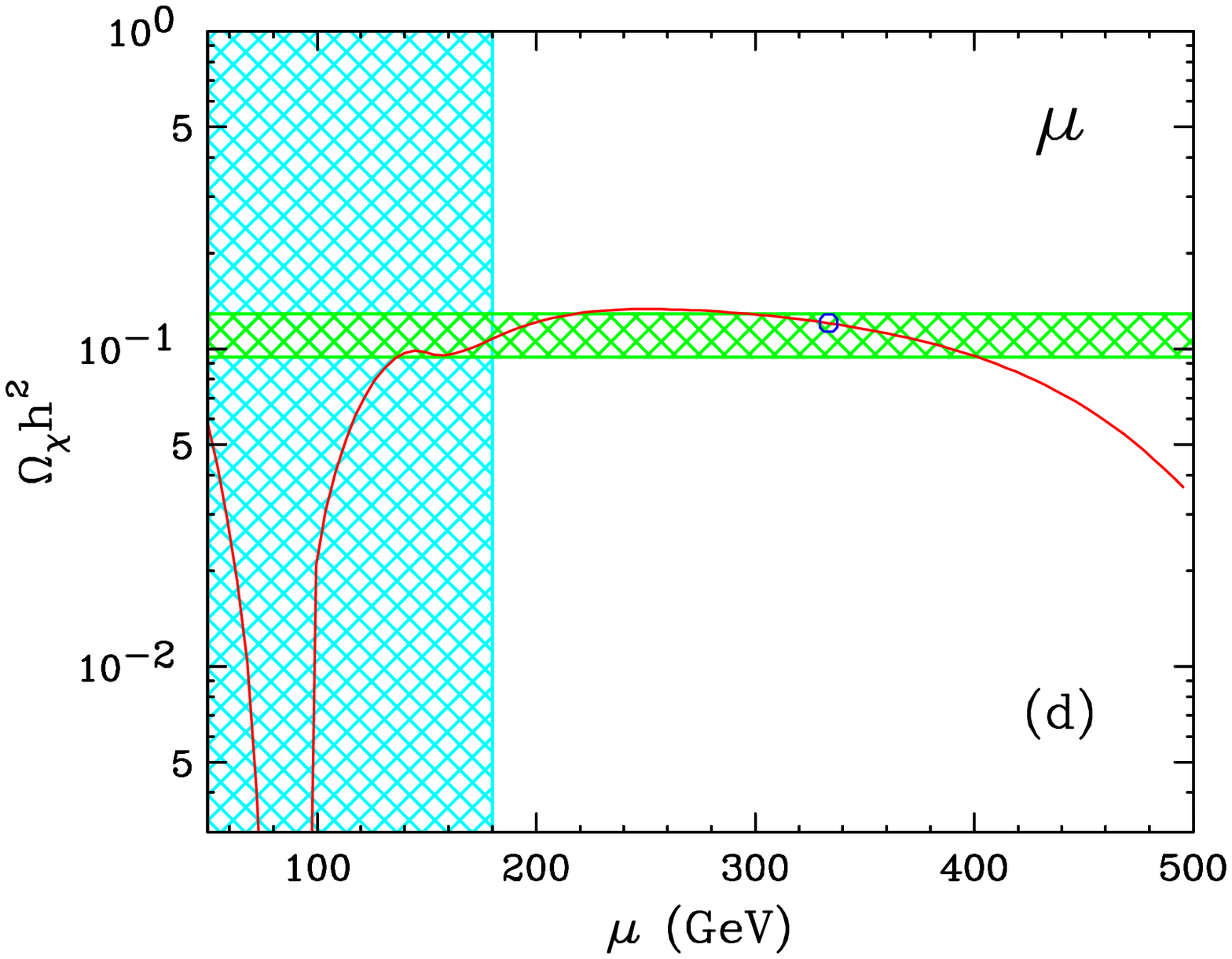}&
\includegraphics[width=50mm]{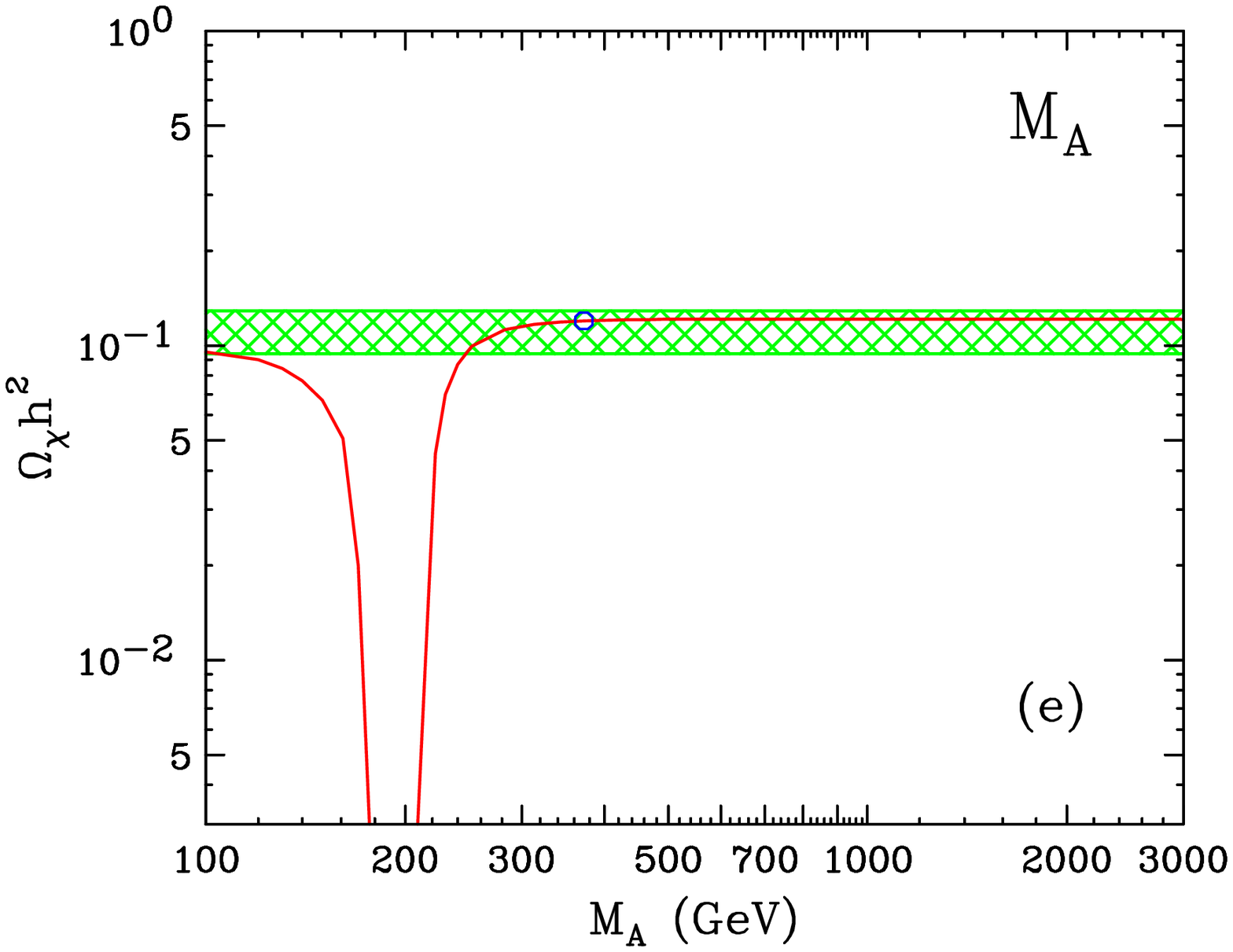}&
\includegraphics[width=50mm]{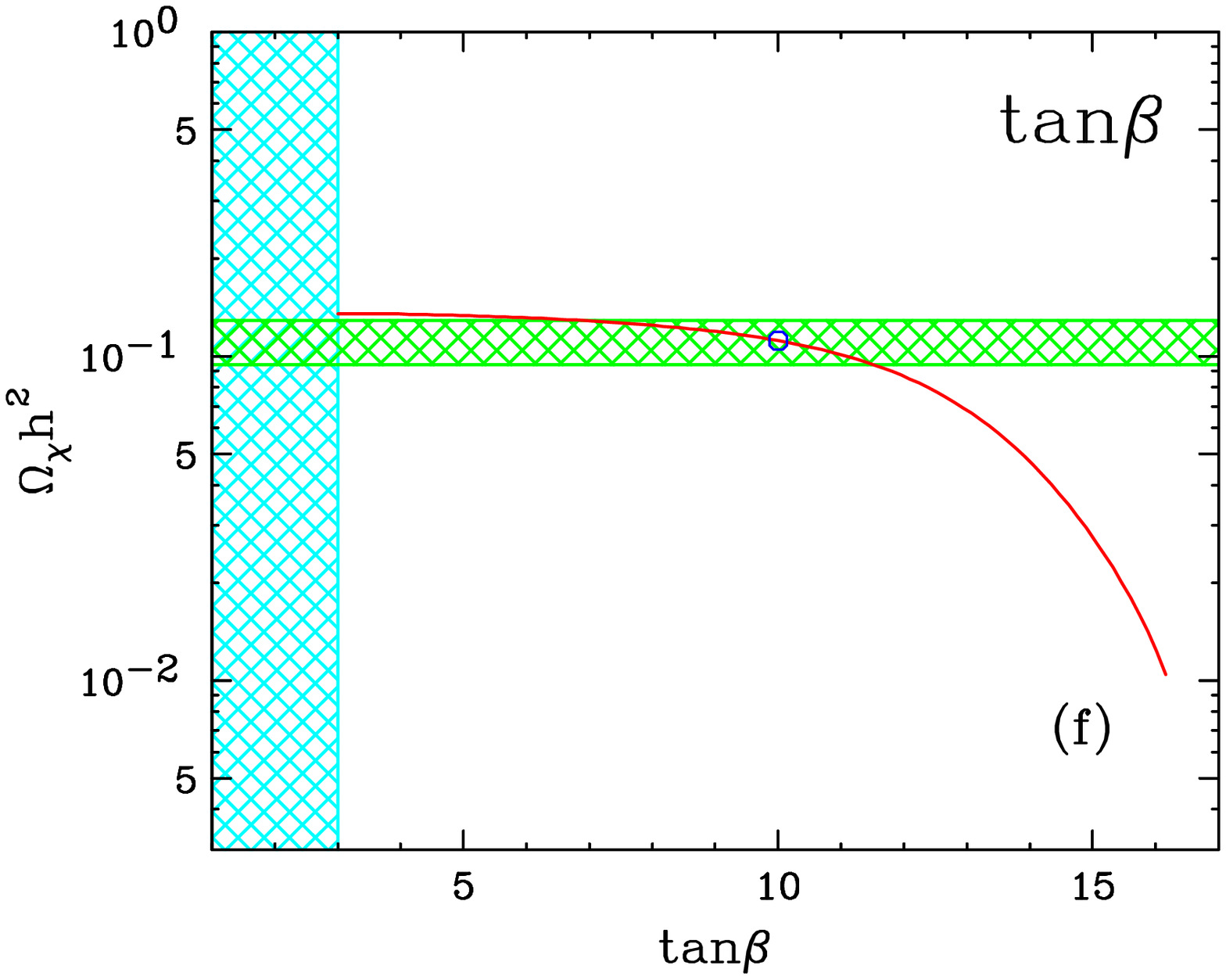}
\end{tabular}

\caption{Effect on relic density of varying relevant SUSY mass parameters for point B'.  
The horizontal (green-shaded) region denotes 
the $2\sigma$ WMAP limits on the dark matter relic density.  The red line shows 
the variation of the relic density as a function of the corresponding SUSY parameter. 
The vertical (blue-shaded) bands denote parameter regions currently ruled out by experiment.
The blue dot in each plot denotes the nominal value for the corresponding parameter at point B'.  Taken from Ref.~\cite{Birkedal:2005jq}.}
\label{fig:Bpvary}
\end{figure}

We first estimate the sensitivity of $\Omega_\chi h^2$
to the various SUSY parameters, and then we determine the precision with which they
can be measured at colliders. In Fig.~\ref{fig:Bpvary} we show the sensitivity of the
dark matter relic density to 6 relevant MSSM parameters. In each panel, the green region 
denotes the $2\sigma$ WMAP limits on $\Omega_\chi h^2$ and the red line shows the 
variation of the relic density as a function of the corresponding parameter.
The vertical (blue-shaded) bands denote parameter regions currently ruled out by experiment.
The blue dot in each panel denotes the nominal value for the corresponding parameter at point B'.

The importance of the different SUSY parameters for the determination of the neutralino 
relic density can be judged from the slope of the lines in Fig.~\ref{fig:Bpvary}. 
If $\Omega_\chi h^2$ is insensitive to a given parameter, the corresponding line 
will be flat.
Conversely, if the relic density is particularly sensitiive to some SUSY parameter, the
slope of the corresponding variation curve will be very steep.
Interestingly enough, Fig.~\ref{fig:Bpvary} also shows that even for one given parameter, there are regions 
where this parameter can be important (for example $M_{\tilde b_R}$ at low masses, where 
coannihilation processes with bottom squarks become important), as well as regions 
where this parameter has very little impact on the relic density (e.g.~$M_{\tilde b_R}$
at its nominal value). Therefore, in estimating the uncertainty on $\Omega_\chi h^2$,
collider data on otherwise ``irrelevant'' parameters can be very important.
In the absence of any information about the value of a given parameter, one should 
let it vary within the whole allowed range, which may encompass values of the
parameter for which it becomes relevant.

The behavior of some of the lines in Fig.~\ref{fig:Bpvary} can be understood as follows.
At point B' the lightest neutralino $\tilde\chi^0_1$ is mostly Bino, hence one 
would expect that its relic density will be sensitive to the Bino mass parameter $M_1$.
Indeed, this is confirmed by Fig.~\ref{fig:Bpvary}(a). For small $M_1$, we observe
enhanced sensitivity near the $Z$ and Higgs pole regions ($2M_{\tilde\chi}\sim M_Z$ 
and $2M_{\tilde\chi}\sim M_h$). For large values of $M_1$ we see a significant variation 
again, this time because the neutralino LSP becomes more and more degenerate with the 
sleptons, and its relic density is depleted due to coannihilation processes.
The analysis of Figs.~\ref{fig:Bpvary}(b) and \ref{fig:Bpvary}(c) is very similar:
the sfermions are irrelevant, if they are heavy, but may become very important if 
they are sufficiently light to induce coannihilations.  For explanations of the behavior 
of the other lines, please refer to Ref.~\cite{Birkedal:2005jq}.

\begin{figure}
  \includegraphics[width=75mm]{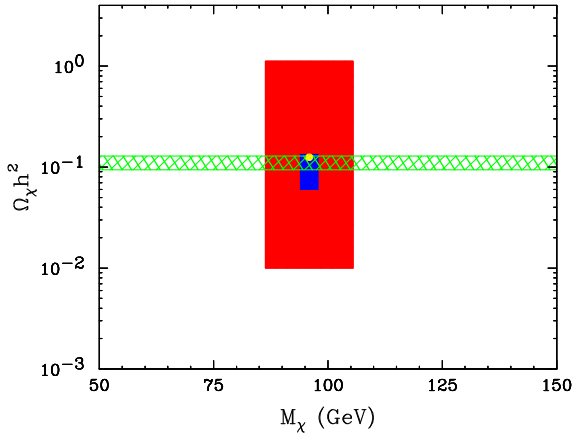}
  \caption{Accuracy of WMAP (horizontal green shaded region), LHC (outer red rectangle) 
and ILC (inner blue rectangle) in determining $M_\chi$, the mass of the 
lightest neutralino, and its relic density $\Omega_\chi h^2$. 
The yellow dot denotes the actual values of $M_\chi$ and 
$\Omega_\chi h^2$ for point B'.  Taken from Ref.~\cite{Birkedal:2005jq} and Ref.~\cite{Birkedal:2004ALCPG}.} 
\label{fig:QUplot}
\end{figure}

Having determined the correlations between the weak-scale SUSY parameters
and the relic abundance of neutralinos, it is now straightforward to estimate
the uncertainty in $\Omega_\chi h^2$ after measurements at different colliders. 
The result is shown in Fig.~\ref{fig:QUplot}, where the outer red 
(inner blue) rectangle indicates the expected uncertainty at the LHC (ILC)
with respect to the mass $M_\chi$ and relic density $\Omega_\chi h^2$
of the lightest neutralino. The yellow dot denotes the actual 
values of $M_\chi$ and $\Omega_\chi h^2$ for point B' and the 
horizontal green shaded region is the current measurement, Eqn.~\ref{wmap}.

In arriving at this result, we made the following assumptions about measurements at the LHC. We assume measurements of the 
$\tilde\chi^\pm_1$, $\tilde\chi^0_2$ and
$\tilde\chi^0_1$ masses at the level of 10\%. The precision on left-handed squark masses
should be no better than the precision on gaugino masses, 
but we have assumed 10\% again. We have conservatively assumed that no slepton or right-handed squark information will be available.
Finally, in terms of Higgs bosons, we expect a detection only of the lightest 
(Standard Model-like) Higgs boson and the absence of a heavy Higgs boson 
signal will simply place the bound $M_A\geq200$ GeV. 

Our assumptions about the corresponding precision at a $500$ GeV ILC were the following.
Since superpartners need to be pair-produced, we take all sparticles lighter 
than 250 GeV to be observable, and their masses can be measured to within 2\%.
This includes the same chargino-neutralino states as in the case of LHC, plus 
all sleptons.

From Fig.~\ref{fig:QUplot} we see that with the assumptions above, the LHC (scheduled to 
turn on in 2007) is not competitive with the current state of the art 
determination of the relic density from cosmology. Nevertheless, it will
bound the relic density from above and below, and may provide the first hint
on whether the dark matter candidate being discovered at colliders is 
indeed the dark matter of cosmology. The ILC will fare much better, and will 
achieve a precision rivalling that of the cosmological determinations.


\begin{figure}
\begin{tabular}{ccc}
  \includegraphics[width=50mm]{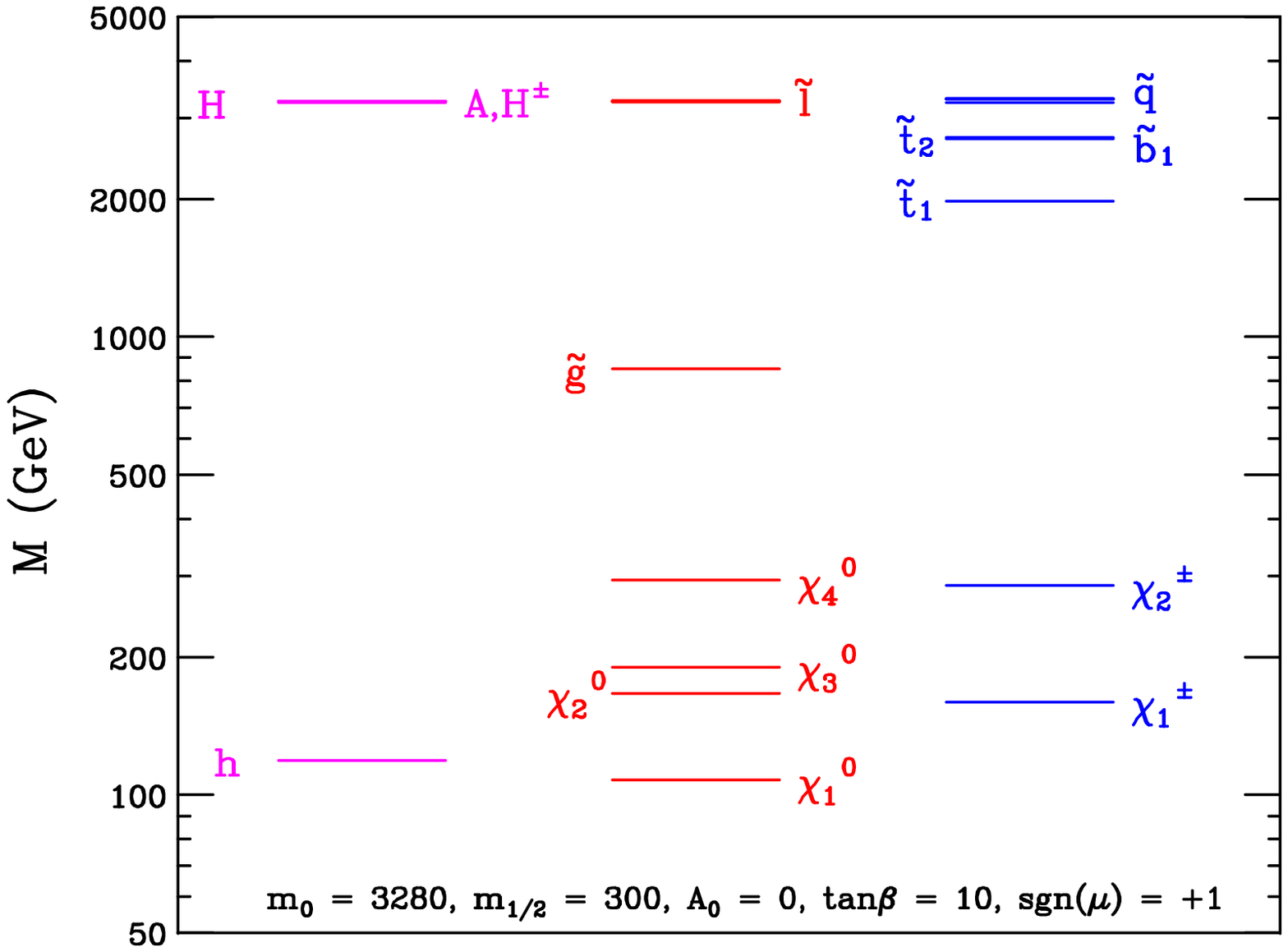}&
  \includegraphics[width=50mm]{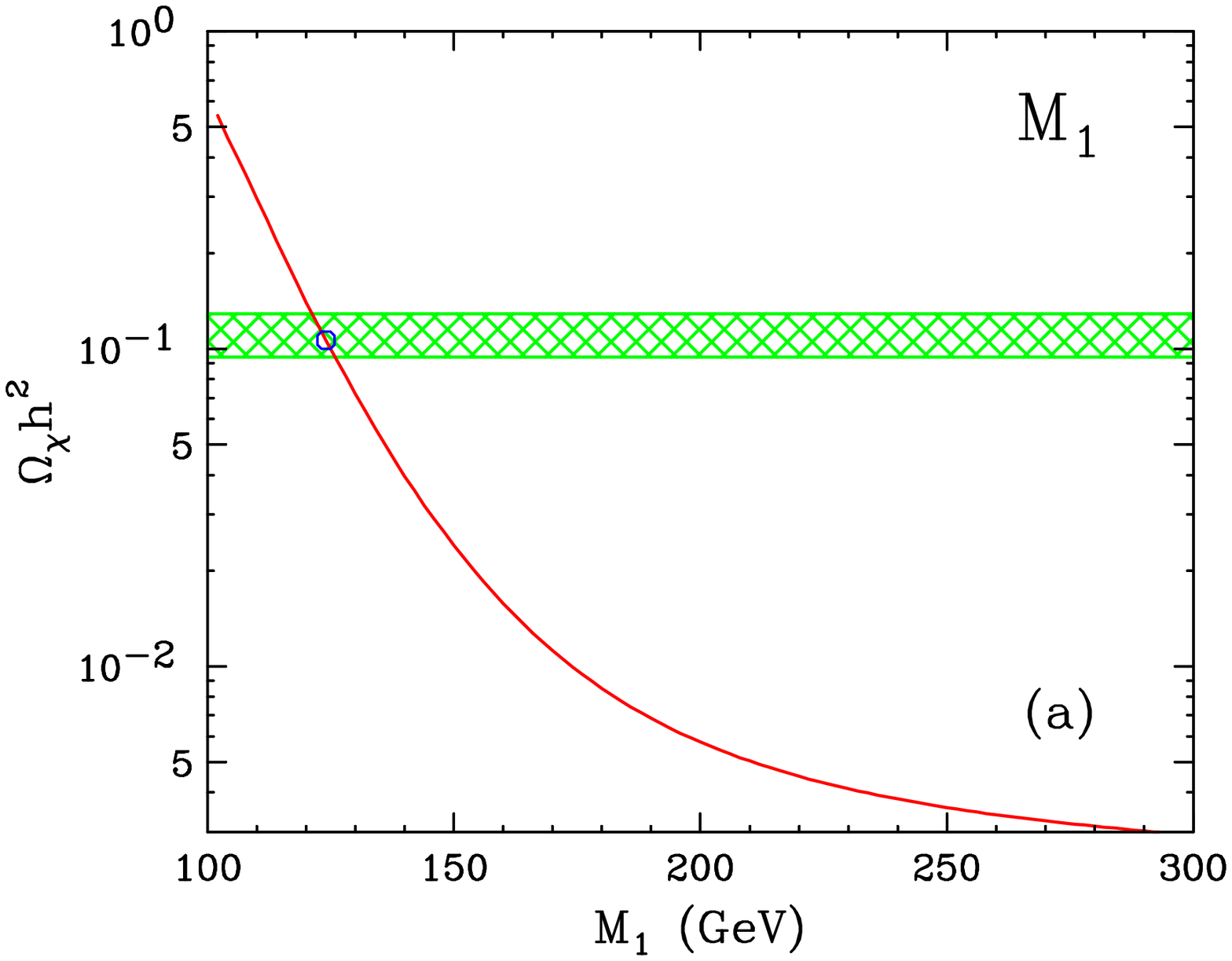}&
  \includegraphics[width=50mm]{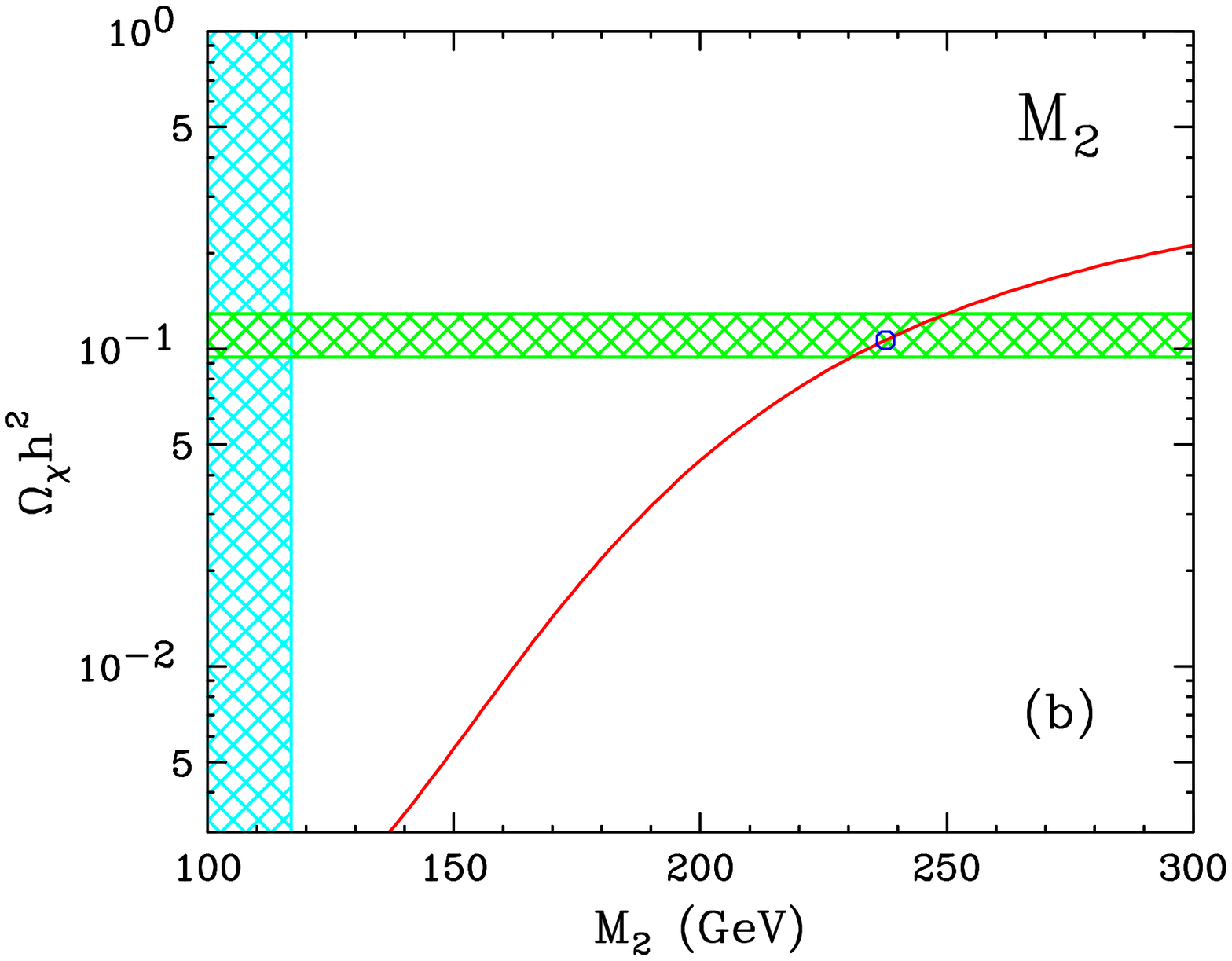}\\
  \includegraphics[width=50mm]{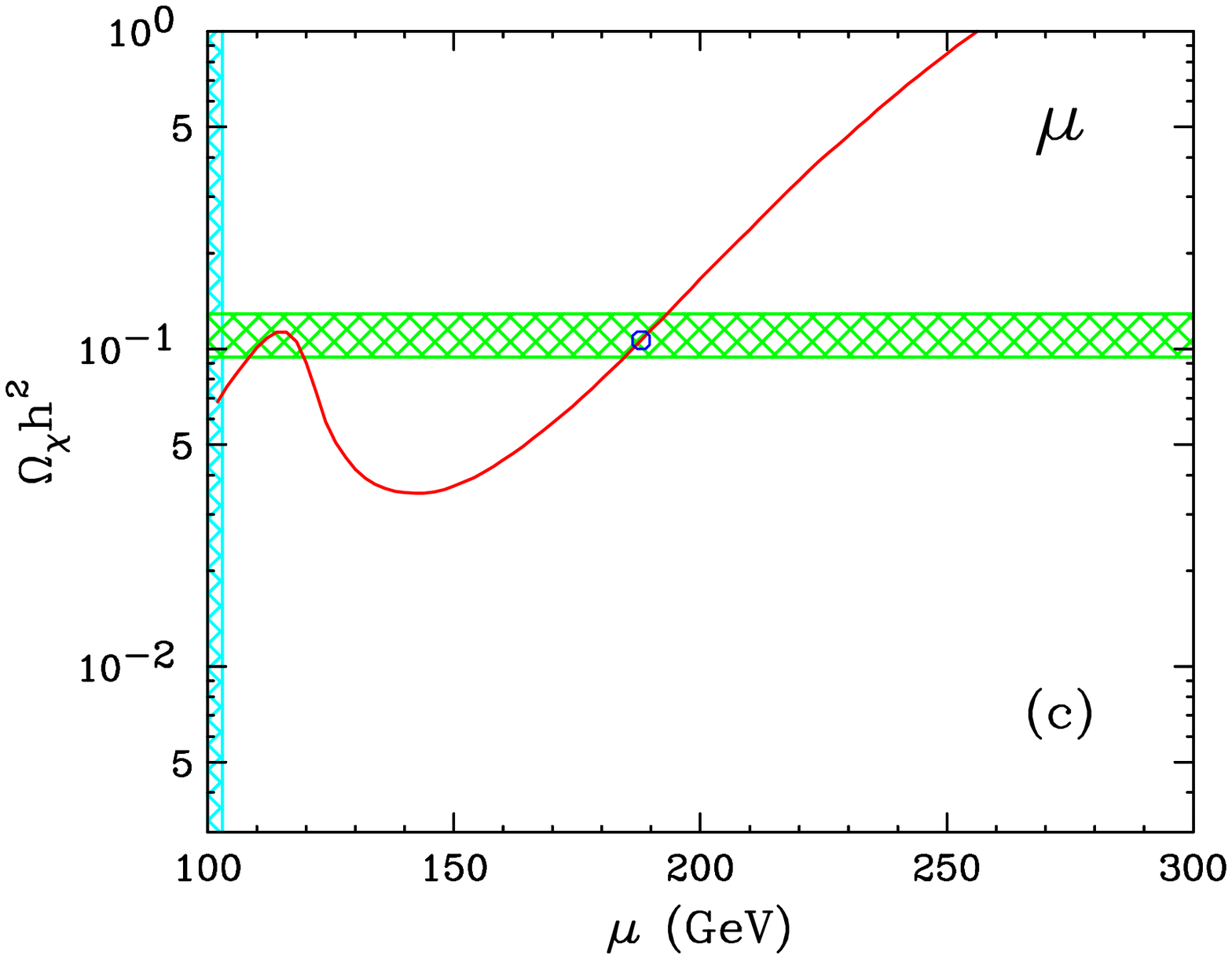}&
  \includegraphics[width=50mm]{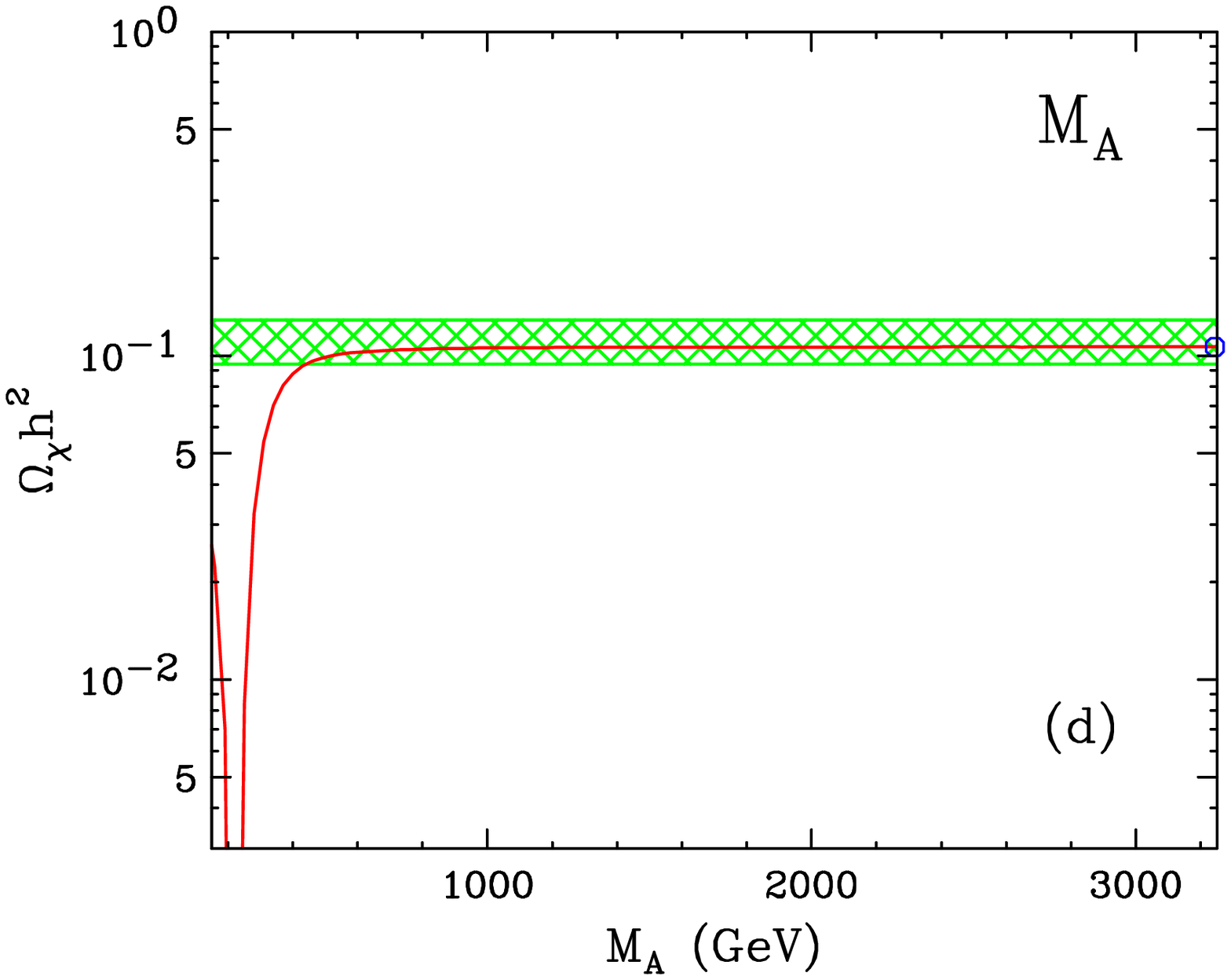}&
  \includegraphics[width=50mm]{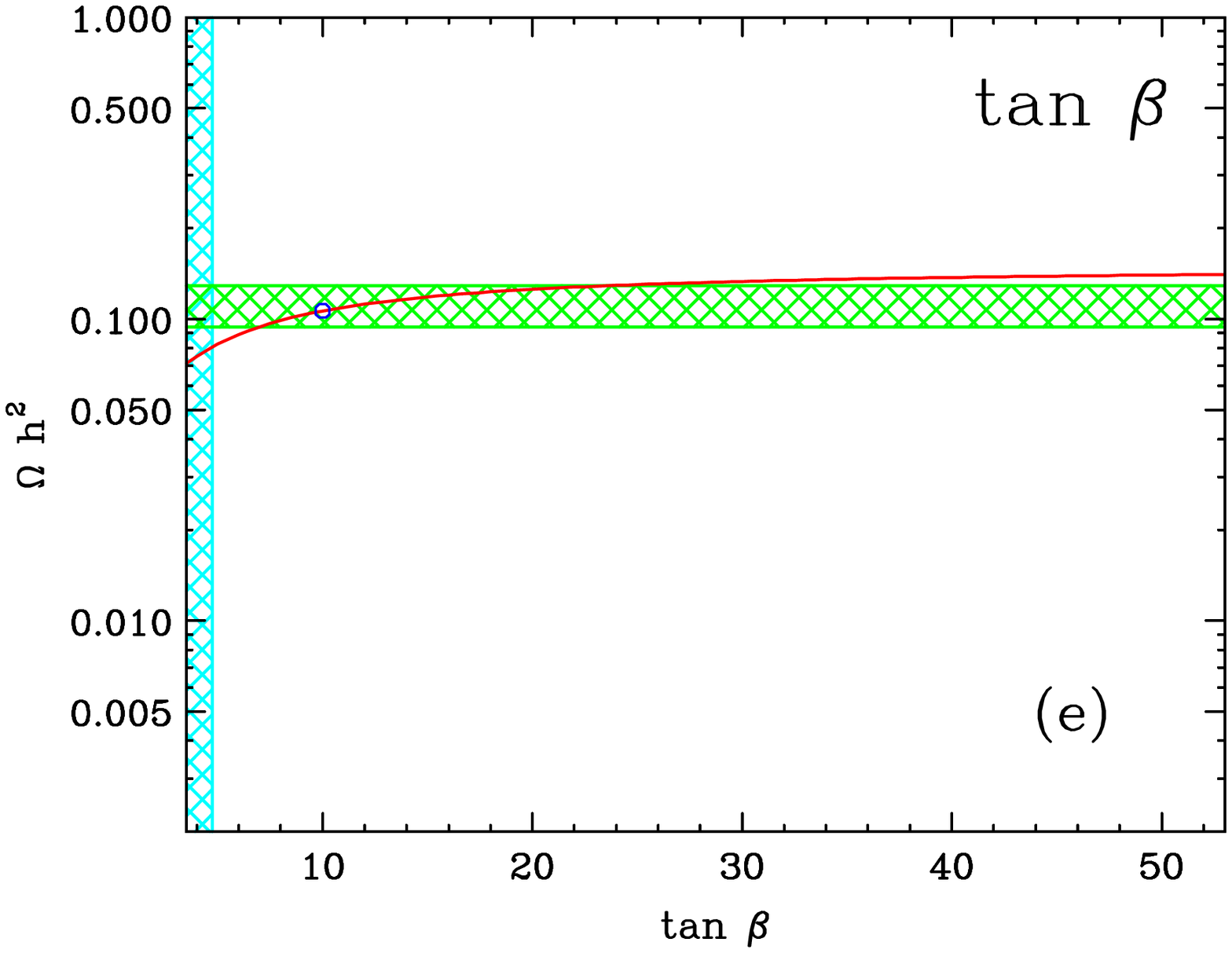}
\end{tabular}
  \caption{{\it Upper left panel:} Mass spectrum of point LCC2 from the focus point region.  {\it Other panels:} Same as Fig.~\ref{fig:Bpvary}, but for point LCC2.  Taken from Ref.~\cite{Birkedal:2005jq}.}
\label{fig:LCC2vary}
\end{figure}

Turning to point LCC2, we show in Fig.~\ref{fig:LCC2vary} the spectrum of particle
masses.  We also show
the analogous variation of the relic density as a function of 5 relevant parameters.
Here the squark and slepton masses are heavy and have little impact on 
the actual $\Omega_\chi h^2$. For this point, the lightest neutralino is mostly 
Bino, but with a non-negligible higgsino component. 

Fig.~\ref{fig:LCC2vary}(a) shows the dependence of $\Omega_\chi h^2$
on the Bino mass parameter $M_1$. As expected, lowering $M_1$ 
increases the Bino component of the LSP, thus suppressing 
$\sigma_{\rm an}$ and increasing $\Omega_\chi h^2$. 
Fig.~\ref{fig:LCC2vary}(b) exhibits complementary behavior: the $M_2$  parameter controls the wino fraction 
of the LSP, and small values of $M_2$ lead to wino-like dark matter, which 
has a large annihilation rate and therefore smaller relic abundance\footnote{Neutralino dark matter with significant wino content has been studied in the contexts of scenarios with non-universal gaugino masses~\cite{Mizuta:1992ja}, string-derived supergravity~\cite{Birkedal-Hansen:2001is} and anomaly-mediated supersymmetry breaking~\cite{Moroi:1999zb}.}.  Again, see Ref.~\cite{Birkedal:2005jq} for further details.

These results will be combined with the outcome of a comprehensive simulation study,
including detailed detector simulation, on the expected experimental precision 
at a $500$ GeV ILC for point LCC2. The final product will be the analogue of 
Fig.~\ref{fig:QUplot} for point LCC2. For further details on the current status of this analysis, see~\cite{Gray:2005ci}. 

\section{Measuring Sleptons at the LHC}

As discussed in the previous section, slepton masses are among the key parameters 
in determining whether $\tilde\chi^0_1$ is a good dark matter candidate~\cite{Drees:1992am}.  Without a collider measurement of the slepton mass, there may be a significant uncertainty 
in the relic abundance calculation.  This uncertainty results because the slepton mass should then be allowed to vary within 
the whole experimentally allowed range.

The importance of slepton discovery is two-fold.  First, supersymmetry predicts a superpartner for every standard model particle.  Therefore, the discovery of the superpartners of the leptons is an important step in verifying supersymmetry.  Second, knowledge of slepton masses is {\it always} important for an accurate determination of the relic abundance of $\tilde\chi_1^0$.  

Here we show that the LHC will indeed have sensitivity to mSUGRA slepton masses, even in the case of heavy sleptons~\cite{Birkedal:2005cm}.  We show that the difference between real and virtual sleptons can be clearly seen.  We also show that in case of virtual sleptons, one can frequently limit the allowed range of their masses, which is equivalent to a rough indirect slepton mass measurement.

Slepton studies at the LHC are challenging.  Direct production of sleptons suffers from large backgrounds, mostly due to $W^+ W^-$ and $t \bar{t}$ production~\cite{Andreev:2004qq}; the methods for slepton mass determination used at the linear collider are not applicable here.  Fortunately, sleptons are produced in sizable quantities at the LHC through cascade decays.  These events can be easily triggered on and separated from the standard model backgrounds.  A common hierarchy in supersymmetric models is $|M_1 | < |M_2 | < |\mu |$.  In that case, sleptons affect the decay $\tilde\chi_2^0 \rightarrow \ell^\pm \ell^\mp \tilde\chi_1^0$.  The resulting dilepton distribution, in principle, contains information about the slepton mass $m_{\tilde{\ell}}$.  This situation is complicated by the fact that $\tilde\chi_2^0$ can also decay through a real or virtual $Z$: $\tilde\chi_2^0 \rightarrow Z^{\left(*\right)} \tilde\chi_1^0 \rightarrow \ell^\pm \ell^\mp \tilde\chi_1^0$.

\begin{figure}
  \includegraphics[width=75mm]{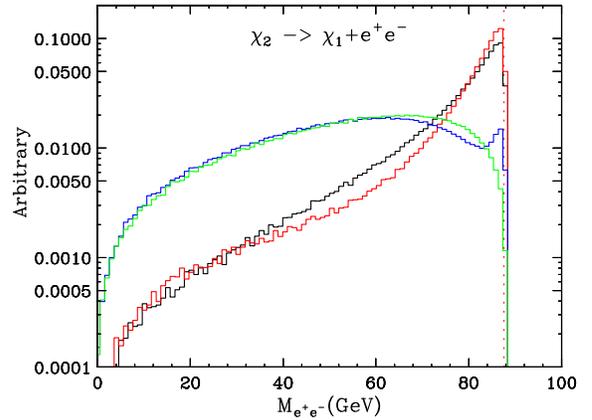}
  \caption{Comparison of $M_{e^+ e^-}$ distributions for different selectron masses.  We only consider the diagrams mediated by $Z$ and $\tilde{e}_R$.  All parameters are held fixed except for the $\tilde{e}_R$ mass.  The (green, blue, red, black) line is for a ($300$, $500$, $1000$ GeV, and $\infty$) mass selectron, respectively.  The neutralino masses are kept constant, and their difference is $88$ GeV.  Taken from Ref.~\cite{Birkedal:2005cm}.} 
\label{fig:InvMassDists}
\end{figure}

What is the observable in these events that is sensitive to the slepton mass?  In this analysis we will consider the dilepton invariant mass distribution, $m_{\ell \ell}$.  The endpoint of $m_{\ell \ell}$ contains information about the masses of the real particles involved in the decay~\cite{Hinchliffe:1996iu}.  But there is more information contained in the $m_{\ell\ell}$ distribution than just in the endpoint~\cite{Birkedal:2005cm}.  One would expect the shapes of the $Z$ and $\tilde{\ell}$ mediated distributions to be different.  Furthermore, the shape of the total decay distribution (including both $Z$ and $\tilde{\ell}$ contributions) should change as a function of the slepton mass.  Indeed this is the case, as illustrated in Figure~\ref{fig:InvMassDists}.  Here we show the dilepton invariant mass distribution resulting from the interference of the $Z$ and $\tilde{e}_R$-mediated diagrams.  The kinematic endpoint is kept fixed; this illustrates that endpoint analyses are largely insensitive to the slepton mass.  In all four cases, the slepton is virtual, but we see a clear difference in the shape of the distribution.  In the case of two body decay through a real slepton, the $m_{\ell\ell}$ distribution will be triangular, as required by phase space.  It is also clear that the $m_{\ell\ell}$ distribution changes significantly as a function of slepton mass

How does one use this fact to statistically differentiate between distributions coming from sleptons of different mass?  The Kolmogorov-Smirnov (K-S) test provides a statistical procedure for differentiating between such distributions~\cite{Frodesen:1979fy}.  The K-S test calculates the maximum cumulative deviation between two unit-normalized distributions.  It then translates this number into a confidence level with which it can be excluded that the two distributions come from the same underlying distribution.  So, the K-S test cannot tell that two distributions came from the same underlying distribution, but it {\it can} tell that two distributions {\it did not} come from the same underlying distribution.  We use this test to explicitly investigate the ability of the LHC to determine slepton masses in mSUGRA.

\begin{figure}[t]
  \includegraphics[angle=270, width = 75mm]{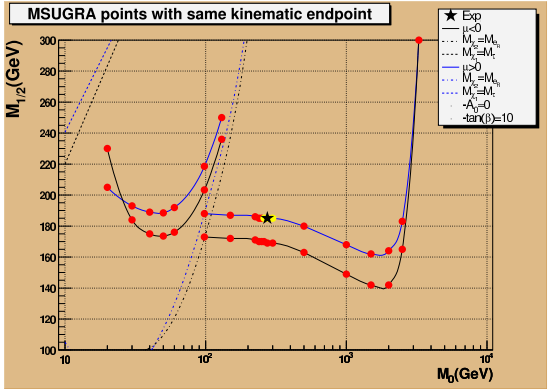}
  \caption{Slepton mass determination in mSUGRA with $A_0 = 0$ and $\tan\beta=10$.  The plot shows the effect on the mSUGRA parameter space of fixing the dilepton kinematic endpoint of the $\chi_{2}^{0} \rightarrow e^+ e^- \chi_{1}^{0}$ decay to be $m_{\ell \ell,max} = 59$ GeV.  It also shows the ability of the K-S test to bound the mass of the lightest selectron for light virtual selectrons.  Details are explained in the text and Ref.~\cite{Birkedal:2005cm}.} 
\label{fig:DistPlots}
\end{figure}

We now demonstrate a slepton mass measurement at the LHC in mSUGRA parameter space.  We select $1000$ events that contain the decay of $\tilde\chi_{2}^{0} \rightarrow e^+ e^- \tilde\chi_{1}^{0}$.  We imagine that the LHC experiments have observed the dilepton mass distribution and have measured a kinematic endpoint at $59$ GeV.  The points with this same kinematic endpoint are the solid lines (blue for $\mu >0$, black for $\mu<0$) in Figure~\ref{fig:DistPlots}.  The dashed lines in the upper left-hand corner indicate where $m_{\tilde\chi_{1}^{0}} = m_{\tilde{\tau}_1}$.  The dashed lines running through the middle of the plot indicate where $m_{\tilde{e}_R} = m_{\tilde\chi_{2}^{0}}$.  This is where the slepton-mediated neutralino decays changes from being three body ($\tilde{e}_R$ is virtual to the right of these lines) to two body ($\tilde{e}_R$ is real to the left of these lines).

In Figure~\ref{fig:DistPlots}, we also display the power of the K-S test in identifying the mass range of the slepton.  We have taken one point (denoted by the black star) as resulting from a possible experimental measurement involving $1000$ signal events.  Then we have compared this 'experimental data' with template distributions (denoted by the red and yellow dots).  The red dots indicate points that the K-S test can identify as not coming from the same distribution as our 'experimental data' at the $95\%$ confidence level.  The yellow dots denote points that could not be excluded at the $95\%$ confidence level.  We can rule out at the $95\%$ confidence level almost all of the example template points except for the ones with the same value of $\mu$, but with almost identical values of $M_0$.  Thus, the mass of the slepton can be determined with some accuracy even though it is not produced as a real particle.  Analysis of experimental points with other values of the slepton mass can be found in Refs.~\cite{Birkedal:2005cm,UsPreprint}.

\section{CONCLUSION}
In this talk, we have introduced techniques and concepts for measuring dark matter at colliders.  We have motivated why this should be possible and also why it is absolutely necessary to validate any claim of a solution to the dark matter puzzle.  We have discussed a challenging model-independent signature of dark matter.  We have also investigated the prospects for making precise measurements of cosmologically relevant parameters in a model-dependent approach.  Finally, we discussed a new technique with implications for dark matter determinations at a collider that allows slepton masses to be measured at the LHC, even for heavy virtual sleptons.


\begin{theacknowledgments}
  The author gratefully acknowledges the help and collaboration of K. Matchev, M. Perelstein, R.C. Group, J. Alexander,K. Ecklund, L. Fields, R.C. Gray, D. Hertz, C.D. Jones, and J. Pivarski.  The author also wished to thank the organizers of PASCOS05 for funding support and a warm invitation.  This work was supported by a US DoE OJI award under grant DE-FG02-97ER41029
\end{theacknowledgments}






\end{document}